# Temperature evolution of quasiparticle dispersion dynamics in semimetallic $1T$-TiTe$_2$ via high-resolution angle-resolved photoemission spectroscopy and ultrafast optical pump-probe spectroscopy

Shuang-Xing Zhu,[1] Chen Zhang,[1] Qi-Yi Wu,[1] Xiao-Fang Tang,[1] Hao Liu,[1] Zi-Teng Liu,[1] Yang Luo,[1] Jiao-Jiao Song,[1] Fan-Ying Wu,[1] Yin-Zou Zhao,[1] Shu-Yu Liu,[1] Tian Le,[2] Xin Lu,[2] He Ma,[3] Kai-Hui Liu,[3] Ya-Hua Yuan,[1] Han Huang,[1] Jun He,[1] H. Y. Liu,[4] Yu-Xia Duan,[1] and Jian-Qiao Meng[1, 5, *]

[1]*School of Physics and Electronics, Central South University, Changsha 410083, Hunan, China*
[2]*Center for Correlated Matter and Department of Physics, Zhejiang University, Hangzhou 310058, China*
[3]*State Key Laboratory for Mesoscopic Physics, Collaborative Innovation Center of Quantum Matter, School of Physics, Peking University, Beijing 100871, China*
[4]*Beijing Academy of Quantum Information Sciences, Beijing 100085, China*
[5]*Synergetic Innovation Center for Quantum Effects and Applications (SICQEA), Hunan Normal University, Changsha 410081, China*
(Dated: Saturday 6$^{\text{th}}$ March, 2021)

High-resolution angle-resolved photoemission spectroscopy and ultrafast optical pump-probe spectroscopy were used to study semimetallic $1T$- TiTe$_2$ quasiparticle dispersion and dynamics. A kink and a flat band, having the same energy scale and temperature-dependent behaviors along the $\overline{\Gamma}$-$\overline{M}$ direction, were detected. Both manifested at low temperatures but blurred as temperature increased. The kink was formed by an electron-phonon coupling. And the localized flat band might be closely related to an electron-phonon coupling. Ultrafast optical spectroscopy identified multiple distinct time scales in the 10-300 K range. Quantitative analysis of the fastest decay process evidenced a significant lifetime temperature dependence at high temperatures, while this starts to change slowly below $\sim$ 100 K where an anomalous Hall coefficient occurred. At low temperature, a coherent $A_{1}g$ phonon mode with a frequency of $\sim$ 4.36 THz was extracted. Frequency temperature dependence suggests that phonon hardening occurs as temperature falls and anharmonic effects can explain it. Frequency fluence dependence indicates that the phonons soften as fluence increases.

## I. INTRODUCTION

Layered transition-metal dichalcogenides (TMDCs) with the chemical formula $TX_2$ ($T$: transition metal; $X$: chalcogen) provide ideal platforms for experimentation due to their quasi-two-dimensional structures and rich physical properties [1–4]. Their $1T$ and $2H$ structures are among the most widely studied. They consist of $X$-$T$-$X$ sandwiches held together by relatively weak "van der Waals" forces between layers [5]. Sample preparation technology has evolved them into valuable materials for studying thin films including monolayer materials [6–8]. Two-dimensional (2D) TMDC materials have several excellent qualities. They are widely used to construct high-performance field-effect and photoelectric devices [9–13]. Two-dimensional metallic TMDC materials such as TaS$_2$ and TiSe$_2$ possess superconductive, charge-density wave (CDW), and several other phases. These qualities provide an excellent research platform for studying various phenomena and the laws of condensed-matter physics [14, 15].

$1T$-TiTe$_2$ is a quasi-2D model material. Its spectral line shapes are consistent with a Fermi-liquid scenario [16–22]. Electron-phonon ($e$-$ph$) coupling was considered when analyzing Ti 3$d$ band spectral line shapes [19–22]. Angle-resolved photoemission spectroscopy (ARPES) measurements have directly observed a weak $e$-$ph$ coupling-induced kink with an energy of $\sim$ 18 meV below the Fermi energy ($E_F$) [23]. Raman scattering revealed two phonon modes: one at $\sim$105 cm$^{-1}$ ($E_g$) and a second at $\sim$143 cm$^{-1}$ ($A_{1g}$) [24, 25]. A CDW phase transition has been found in many TMDC materials [26, 27] but not in bulk TiTe$_2$. It was thought that weak $e$-$ph$ coupling cannot lead to superconductivity or CDW instability.

CDW order was recently observed in single-layer TiTe$_2$ with a transition temperature of 92$\pm$3 K [28], and also in monolayer and multilayer epitaxial TiTe$_2$ films at room temperature [29]. In prior ARPES experiments, $1T$-TiTe$_2$ has been found to exhibit typical semimetallic characteristics with a negative indirect band gap of $\sim$ 0.8 eV [18]. Theory suggested that both doping [30] and strain [31] can lead to a topological phase transition. Experimental results suggest that pressure can cause both topological phase and structural transitions [32]. It may even lead to superconductivity [33].

The complex and numerous physical properties of solid materials are known to be closely related to the many-body interactions among charge, lattice, spin, and orbit. When phase transitions occur, specific changes occur in many-body interactions. For example, energy, and evolution of $e$-$ph$ coupling trends, may change with temperature. Exploring the many-body interaction modes and their evolutions with temperature provides a better approach towards understanding the situation.

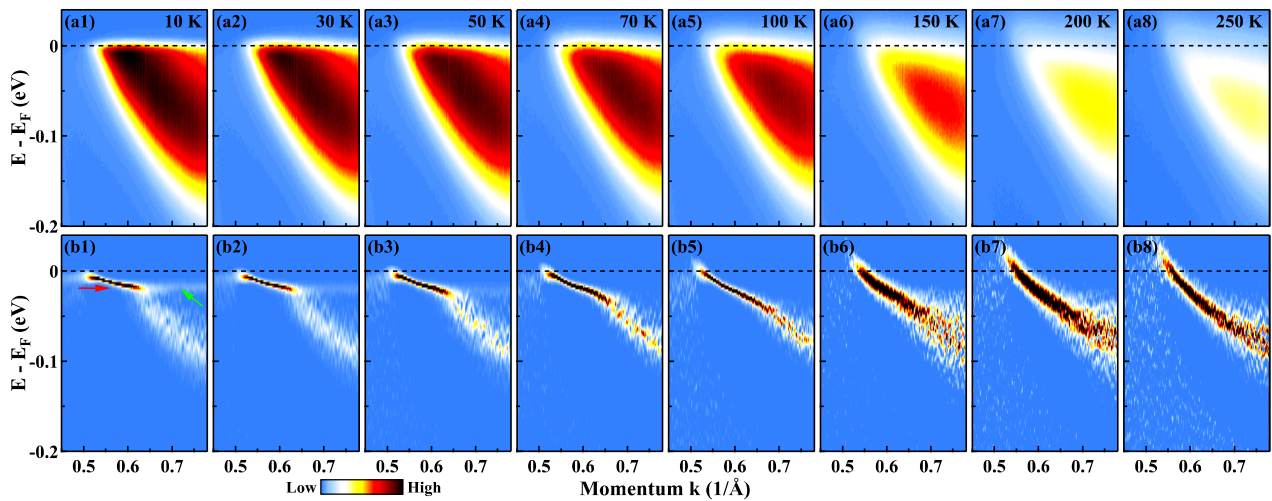

FIG. 1. (Color online) ARPES maps for the 1$T$-TiTe$_2$ Ti 3$d$ band. (**a1-a8**) The band structure plots along the $\overline{\Gamma}$-$\overline{M}$ direction at different temperatures. (**b1-b8**) Second-derivative images to enhance the weak bands corresponding to (**a1-a8**). The red and green arrows mark the kink and flat band, respectively.

ARPES is a powerful tool for studying many-body interactions, as it provides direct information on single-particle spectral functions. Ultrafast optical pump-probe spectroscopy is also useful in studying ultrafast quasiparticle dynamics in the time domain, as transmission/reflectivity changes of the optical probe have been inferred to result from the photoexcited quasiparticles and collective excitations [34, 35].

In this paper, ARPES and ultrafast optical pump-probe spectroscopy are jointly used to study single-crystalline 1$T$-TiTe$_2$ quasiparticle dispersion and dynamics. At 18 meV below $E_F$, ARPES detected a kink induced by $e$-$ph$ coupling and a flat band closely related to the $e$-$ph$ coupling. At low temperatures, a coherent $A_{1g}$ phonon mode with a frequency of ∼4.36 THz was extracted via ultrafast optical pump-probe spectroscopy. As temperature fell, the $A_{1g}$ phonon mode hardened monotonically (blue shift), which can be well explained by an anharmonic phonon model.

## II. SAMPLES AND EXPERIMENTAL MEASUREMENT

High-quality 1$T$-TiTe$_2$ single crystals were grown via a chemical vapor transport method with iodine as the transport agent. Shiny platelike single crystals as large as 10 mm × 10 mm × 0.5 mm were obtained.

High-resolution temperature-dependent ARPES measurements were performed at beamline 5-4 of the Stanford Synchrotron Radiation Lightsource (SSRL) using a Scienta R4000 electron energy analyzer. All samples were cleaved *in situ* and measured in a ultrahigh vacuum with base pressure better than 3×10$^{-11}$ mbar. A 20-eV photon energy beam with an overall energy resolution of ∼7 meV was selected to probe the temperature evolution of $e$-$ph$ coupling in Ti 3$d$ band along the high-symmetry $\overline{\Gamma}$-$\overline{M}$ direction. This photon energy can achieve a very narrow quasiparticle bandwidth with a clear $e$-$ph$ coupling-induced kink structure [23]. Angular resolution was kept at 0.2° for all measurements.

Ultrafast optical pump-probe measurements were carried out with a pulse laser produced by a Yb-based femtosecond (fs) laser oscillator. The pulses had a center wavelength of 800 nm (1.55 eV), a pulse width of ∼ 35 fs, and a low repetition rate of 1 MHz. Pump and probe beams spot diameters on the sample were 160 and 40 $\mu$m, respectively. Pump and probe beams were $s$- and $p$-polarized, respectively. Data were collected from 10 K up to 300 K on freshly cleaved surfaces. All measurements were taken in vacuum (10$^{-6}$ Torr).

## III. ARPES EXPERIMENTAL

Figures. 1(a1)-(a8) show the 1$T$-TiTe$_2$ temperature-dependent energy-momentum images. As the temperature rose, quasiparticle dispersion intensity weakened. No evidence of CDW was observed here, such as a replica of Te 5$p$ bands shifted from $\overline{\Gamma}$ to $\overline{M}$ at low temperatures in single-layer TiTe$_2$ [28]. Shallow Ti 3$d$ band dispersion [23] is not easily extracted using fitting momentum distribution curves (MDCs) or energy distribution curves (EDCs) [16–23] (Fig. S1 of the Supplemental Material [36]). Figures. 1(b1)-(b8) present the quasiparticle dispersion reduced by a second derivative along the energy direction to sharpen the band structures while maintaining main band structure. A kink structure shown by the

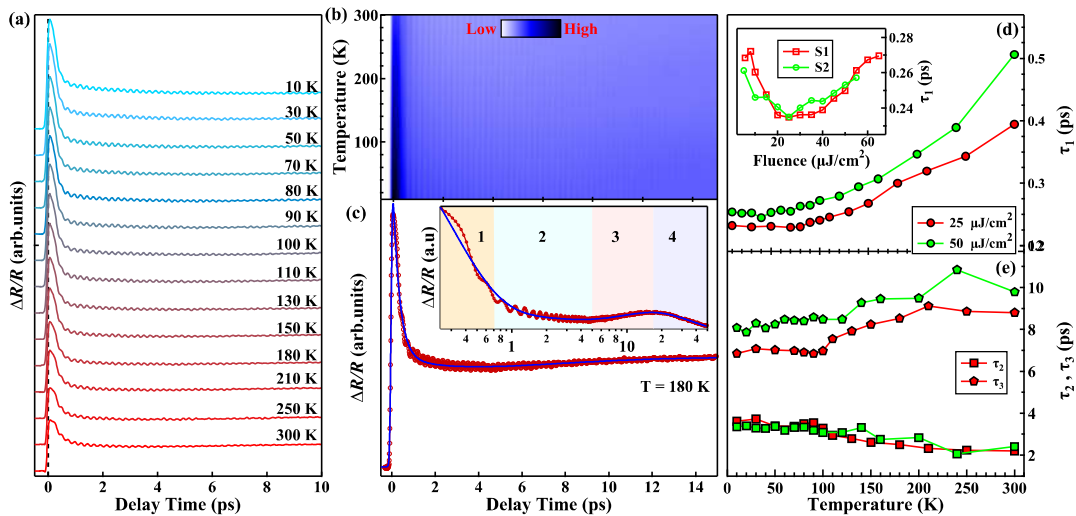

FIG. 2. (Color online) (**a**) $\Delta R/R$ as a function of delay time over temperature range from 10 to 300 K at a pump fluence of $\sim 25\ \mu J/cm^2$. (**b**) 2D pseudocolor map showing $\Delta R/R$ as a function of temperature and delay times. (**c**) The blue line is a typical fitting of $\Delta R/R$ at 180 K using four independent exponential decays. The inset shows four relaxation regions using different colors: Region 1 is the quickest relaxation process in $\sim 0.7$ ps. Region 2 shows a slower relaxation process between $\sim 0.7$ and $\sim 5$ ps. Region 3 shows a second rise occurring between $\sim 5$ and $\sim 17$ ps. Region 4 shows a very slow relaxation process lasting for more than 30 ps. (**d, e**) The extracted lifetimes $\tau_1$, $\tau_2$, and $\tau_3$ as a function of temperature at two different fluences. $\tau_1$, $\tau_2$, and $\tau_3$ represent the relaxation processes 1, 2, and 3 indicated in the inset of (c), respectively. The inset of (d) displays $\tau_1$ as a function of the pump fluence for two different samples (S1 and S2).

red arrow was detected at around 18 meV below the $E_F$ at low temperatures. Previous work has well discussed the kink structure and attributed it to e-ph coupling [23]. The kink blurred as the temperature rose, until becoming indistinguishable at around 70 K.

A flat band near $E_F$ marked by the green arrow was well resolved in the second-derivative figures at low temperatures. Theory did not predict the existence of a flat band in this material. Flat bands are common in TMDC materials, such as TiTe$_2$ [29, 39], TaS$_2$ [40], TaSe$_2$ [40], and VSe$_2$ [41]. They are also present in 2D electron liquid [42] and in strongly correlated materials such as twisted bilayer graphene [43–45] and Bi$_2$Sr$_2$CaCu$_2$O$_{8+\delta}$ [46].

The flat band was thought to result from narrow impurity band emissions [39, 47], e-ph coupling [48], a polaron effect associated with CDW order [29, 39], or e-ph coupling [42]. A CDW-caused polaron effect is unlikely, as no CDW order was observed in bulk TiTe$_2$. The flat band has the same energy as the e-ph coupling-induced kink. Its temperature-dependent behavior is identical and readily detected at low temperatures. The flat band weakens as temperature increases until becoming indiscernible around 70 K. The carrier density of 1$T$-TiTe$_2$ is $\sim 10^{20}$-$10^{21}$ cm$^{-3}$ ($\sim 10^{13}$-$10^{14}$ cm$^{-2}$) [39], which is conducive to polaron formation [42]. e-ph coupling may be crucial to flat band formation. ARPES spectral measurements did not reveal replica bands separated by e-ph coupling mode energy. This may be due to limited energy resolution. We note that our experimental spectra cannot rule out the possibility of an emission from a narrow impurity band. Prior studies of Ti 3$d$ spectral line shapes have suggested impurity contributions at around 14 [22] or 17 meV [19, 21], which is similar to the location of the flat band.

It can also be seen from the figures that with the decrease of temperature, the Fermi momentum shifts to the left. This means that hole pocket volume formed by the Ti 3$d$ band increases as temperatures fall. In addition, the volume of electron pockets formed by Te 5$p$ bands also increases as temperatures fall (Fig. S2 of the Supplemental Material [36]). Both hole and electron pocket volumes change with temperature. This may explain the presence of a Hall coefficient anomaly in TiTe$_2$ [39]. The Hall effect is believed to be composed of electron and hole contributions which nearly offset each other. The Hall effect is sensitive to small changes of carrier concentration in a simple two-carrier model. The band shift in this study suggests a possible explanation for the long standing uncertainty about the Hall coefficient anomalies.

## IV. ULTRAFAST OPTICAL PUMP-PROBE MEASUREMENT

Carrier dynamics in solids are sensitive to many-body interactions among the charge, lattice, spin, and orbital. Ultrafast optical spectroscopy is a potent tool for investigating carrier dynamics. It has been used to study heavy fermions [49, 50], high-temperature superconduc-tivity [51–53], semiconductors [54], topological materials [35, 55, 56], and CDW materials [34]. Ultrafast optical pump-probe measurements were performed on single-crystal 1$T$-TiTe$_2$ to investigate quasiparticle dynamics and phonon modes.

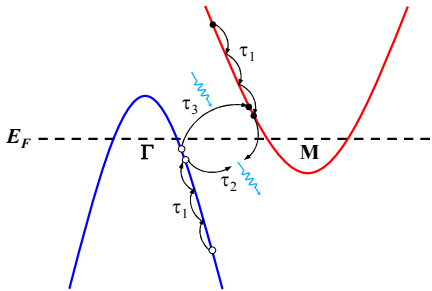

FIG. 3. (Color online) Schematic band structure of semimetal 1$T$-TiTe$_2$ near the $E_F$ along the $\overline{\Gamma}$-$\overline{M}$ direction. $\tau_1$, $\tau_2$, and $\tau_3$ represent $e$-$ph$ thermalization, phonon-assisted interband $e$-$h$ recombination, and reexcited $e$-$h$ pairs, respectively.

Figure. 2(a) presents $\Delta R/R$ signals for 1$T$-TiTe$_2$ at various temperatures at a pump fluence of $\sim 25$ $\mu$J/cm$^2$. Upon photoexcitation, $\Delta R/R$ signal changes instantaneously. It is then followed by a long recovery process. Reflectivity signal peak intensity increases as temperature decreases. The $\Delta R/R$ signal oscillates significantly at all measured temperatures, becoming more pronounced as temperature drops. Figure. 2(b) is a 2D pseudo color $\Delta R/R$ mapping image shown as a function of pump-probe delay ($x$ axis) and temperature ($y$ axis).

Nonoscillatory background decay does not change significantly with temperature. Relaxation can be roughly divided into four stages [Fig. 2(c) inset and Fig. S3 of the Supplemental Material [36]]. A fast recovery process ($\tau_1$) occurs within 0.7 ps, followed by a slower component ($\tau_2$). A pervasive second rising ($\tau_3$) occurs within a few picoseconds. This is often found in strongly correlated materials [50–52] and topological insulators [55]. The final relatively slow relaxation ($\tau_4 > 20$ ps) is deemed a phonon-acoustic-phonon interaction.

Quantitative analysis of quasiparticle dynamics was conducted to study temperature and fluence dependence behavior. The solid blue line in Fig. 2(c) suggests that the nonoscillatory response fits well with four exponential decays convoluted with a Gaussian laser pulse, using the expression

$$\frac{R(t)}{R} = \frac{1}{\sqrt{2\pi}w}\exp(-\frac{t^2}{2w^2}) \otimes [\sum_{i=1}^{4} A_i \exp(-\frac{t-t_0}{\tau_i})] + C,$$

where $A_i$ is the amplitude and $\tau_i$ is the relaxation time of the $i$th nonoscillatory signal which describes carrier dynamics. $w$ is incidence pulse temporal duration. $C$ is a constant offset.

Decay times $\tau_1$, $\tau_2$, and $\tau_3$, denoted by red symbols in Figs. 2(d) and 2(e), are a function of temperature. Lifetime changes with temperature at a pump fluence of $\sim 50$ $\mu$J/cm$^2$ are indicated in green. For both fluences, $\tau_1$ shows significant temperature-dependent behavior at high temperature. $\tau_1$ begins changing slowly at around $\sim 100$ K. This is immediately below the temperature when thermoelectric power $\alpha$ changes sign and an anomalous Hall coefficient occurs [39]. Precise temperature-dependent resistivity measurements revealed that the resistivity derivative $d\rho/dT$ has a broad hump at the same temperature scale which starts at $\sim 125$ K and centers $\sim 100$ K (Fig. S4 of the Supplemental Material [36]). The time scales of the initial decay, $\tau_1$, are plotted as a function of pump fluence and appear in the inset of Fig. 2(d). In the instant experimental resolution, the change of $\tau_1$ was negligible in the wide range of pump fluences used here. This suggests that the laser pulse did not cause a significant change in the material during the experiment.

The following were considered in order to determine possible physical origins of the relaxation processes. Decay origins are revealed by analyzing the 1$T$-TiTe$_2$ electronic band structure. Typical semimetallic electronic structures were observed in 1$T$-TiTe$_2$, including multiple hole pockets around $\Gamma$ points and electron pockets around $M(M')$ points [23]. Electron and hole pockets dominate the Fermi surface structure.

Figure. 3 shows the illustration of the band structure of 1$T$-TiTe$_2$ near the $E_F$. In the initial stages photo and thermal carriers rapidly lose excess high energy through $e$-$ph$ interactions ($\tau_1 < 0.7$ ps). Electron-optical-phonon thermalization lifetimes decreased as temperature decreased and saturated at $\sim 0.24$ ps at low temperatures. This robust temperature dependence indicates that the rapid relaxation process, $\tau_1$, is caused by an $e$-$ph$ interaction rather than from electron-electron scattering. At high temperatures, $e$-$ph$ thermalization is generally described by a two-temperature model (TTM)[38], which suggests $e$-$ph$ thermalization time is proportional to the sample temperature. According to the TTM [38, 57], the relaxation rate, $\gamma_T$, of the quasiparticles is given by

$$\gamma_T = \frac{3\hbar\lambda\langle\omega^2\rangle}{\pi k_B T_e}[1 - \frac{\hbar^2\langle\omega^4\rangle}{12\langle\omega^2\rangle k_B^2 T_e T_l} + \cdots],$$

where $T_e$, $T_l$, $\lambda\langle\omega^2\rangle$, $\lambda$, and $\omega$ are electron temperature, lattice temperature, the second moment of the Eliashberg function, $e$-$ph$ coupling constant, and phonon frequency, respectively. $T_e$ is a constant greater than $T_l$ at low temperatures and increases slowly with the increase of $T_l$ at high temperatures [57]. In this way, temperature or energetic differences between $T_e$ and $T_l$ decrease as temperature increases. This results in lengthened $e$-$ph$ thermalization times at high temperatures [Fig. 2(d)].

After fast $e$-$ph$ thermalization takes place, quasiparticles decay to the Fermi edge. $\tau_2$ may arise from a phonon-assisted electron-hole ($e$-$h$) recombination between the conduction and valence bands, which is com-mon in semimetals [58, 59]. Excited quasiparticle decay results in high-frequency boson emission. This can subsequently reexcite e-h pairs and lead to a secondary reflectivity rise. The amplitude, $A_3$, of the secondary rise is negative. It is very weak at low temperature and becomes more significant as temperature increases (Fig. S3 of the Supplemental Material [36]). This means that the more phonons are excited, the stronger of the reexcitation.

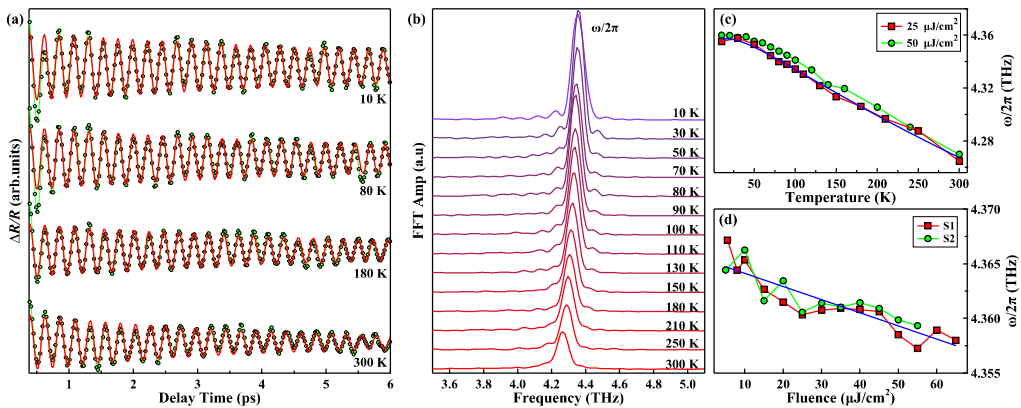

FIG. 4. (Color online) (a) Oscillations after subtracting nonoscillatory background at a pump fluence of $\sim 25$ $\mu J/cm^2$ for four typical temperatures: 10, 80, 180, and 300 K. The fitted results are in red. (b) Fast Fourier transform (FFT) frequency-domain data corresponding to (a). Curves are shifted vertically for clarity. (c) The derived frequency $\omega/2\pi$ as a function of temperature at the two different fluences. The blue line is the fitted curve of the lower fluence data (red squares) using an anharmonic phonon model. (d) Pump fluence dependence of the frequency $\omega/2\pi$ at 10 K for two different samples (S1 and S2). The blue line serves as a guide for the eyes.

As temperature decreases, $\tau_2$ gradually increases and $\tau_3$ gradually decreases [Fig. 2(e)]. This supports the conclusion that $\tau_2$ and $\tau_3$ are induced by the coupling of quasiparticles with phonon excitations, since phonon population depends on temperature. A temperature decrease results in a Ti $3d$ band shift towards the $\overline{\Gamma}$ point. The relative positions of the valence band and the conduction band vary with temperature changes. This will affect e-h recombination and reexcitation.

$\Delta R/R$ signal oscillations were analyzed for collective excitations and even possible CDW phase transitions. The time-domain oscillations for several typical temperatures at a pump fluence of $\sim 25$ $\mu J/cm^2$ appear in Fig. 4(a). Oscillations exist at all measured temperatures up to room temperature. The oscillatory components that exist at room temperature are normally attributed to coherent phonons. The red curves are the fitted results based on this damped oscillation function:

$$(\Delta R(t)/R)_{osc} = Ae^{-\Gamma t}\sin(\omega t + \varphi),$$

where $A$, $\Gamma$, $\omega$, and $\varphi$ are the amplitude, damping rate, frequency, and phase, respectively.

Figure. 4(b) shows the oscillatory components of Fig. 4(a) extracted by a fast Fourier transform (FFT). A distinct terahertz mode was observed at all measured temperatures with a frequency $\omega/2\pi \sim 4.36$ THz at 10 K. This mode of corresponding energy is about 18 meV and is consistent with both the ARPES experiment kink energy [23] and the Raman measurement of the $A_{1g}$ phonon ($\sim 145$ cm$^{-1}$) [25? ]. The oscillation mode appears to be a coherent $A_{1g}$ phonon. The derived frequency is plotted in Fig. 4(c) as a function of temperature. The frequency extracted with a pump fluence of $\sim 50$ $\mu J/cm^2$ is added and represented with green circles. $\omega/2\pi$ becomes nearly constant at low temperatures. For both pump fluences, the $A_{1g}$ phonon mode hardens monotonically with decreasing temperature (blue shift), from $\sim 4.26$ THz at 300 K to $\sim 4.36$ THz at 10 K. Temperature dependence here is well explained by optical phonon anharmonic effects [60, 61]. These are also thought to be important in $1T$-TiTe$_2$ [62].

Figure. 4(d) displays the derived frequency $\omega/2\pi$ at 10 K as a function of pump fluence. , Frequency decreases linearly with increasing excitation fluences at a rate of -0.00012 THz per $\mu J/cm^2$ for the pump fluence used here. $A_{1g}$ phonon hardening below $\sim 8$ GPa has been observed when high hydrostatic pressure was applied [25]. It has been suggested that this hardening is be due to state densities near the $E_F$, which decreases dramatically with hydrostatic pressure [63]. The phonon softening observed here may also closely relate to the density of electronic states. Laser radiation is thought to weaken interatomic bonds, leading to atomic displacement, an increase in the density of states of electrons at $E_F$, and finally to phonon softening [64–67].

## VI. CONCLUSIONS

High-resolution ARPES and ultrafast transient reflectivity measurements have been performed on bulk $1T$-TiTe$_2$. ARPES revealed a kink and a flat band at 18 meV along the $\overline{\Gamma}$-$\overline{M}$ direction. The kink and flat band have both the same energy scale and temperature-dependent behavior. This is observable at low temperatures and blurs as temperatures increase. The kink is formed by e-ph coupling. The e-ph coupling might play an important role in the formation of a localized flat band. In the 10-300 K range, multiple decay processes were detected by ultrafast optical spectroscopy. $\tau_1$, $\tau_2$, and $\tau_3$ are ascribed to carrier cooling through e-ph thermalization, phonon-assisted e-h recombination, and reexcitation, respectively. $\tau_4$ is considered to be phonon-acoustic-phonon interaction. At 10 K, a coherent $A_1g$ phonon mode with a frequency of $\sim$ 4.36 THz was extracted. It was consistent with ARPES and Raman measurements. The temperature dependence of the $A_1g$ mode frequency can be well explained using an anharmonic phonon model.

## VII. ACKNOWLEDGEMENT

We thank Gey-Hong Gweon for his unreserved support. This work was supported by the Innovation-Driven Plan in Central South University (No. 2016CXS032). X. Lu was supported by the National Natural Science Foundation of China (NSFC, Grant No. 11674279).

* Corresponding author: jqmeng@csu.edu.cn
[1] B. Sipos, A. F. Kusmartseva, A. Akrap, H. Berger, L. Forro, and E. Tutis, nature materials **7**, 960 (2008).
[2] K. F. Mak, K. He, J. Shan, and T. F. Heinz, Nat. Nanotechnol **7**, 494-498(2012).
[3] M. N. Ali, J. Xiong, S. Flynn, J. Tao, Q. D. Gibson, L. M. Schoop, T. Liang, N. Haldolaarachchige, M. Hirschberger, N. P. Ong, and R. J. Cava, Nature (London) **514**, 205-208(2014).
[4] M. Bonilla, S. Kolekar, Y. Ma, H. C. Diaz, V. Kalappattil, R. Das, T. Eggers, H. R. Gutierrez, M. Phan, and M. Batzill, Nature Nanotechnology **13**, 289-293(2018).
[5] S. Bocharov, G. Dräger, D. Heumann, A. Šimmunek, and O. Šipr, Phys. Rev. B **58**, 7668 (1998).
[6] C. S. Cucinotta, K. Dolui, H. Pettersson, Q. M. Ramasse, E. Long, S. E. OBrian, V. Nicolosi, and S. Sanvito, J. Phys. Chem. C **119**, 15707 (2015).
[7] B. Singh, C. H. Hsu, W.F . Tsai, V. M. Pereira, and H. Lin, Phys. Rev. B **95**, 245136 (2017)
[8] K. H. Park, J. Choi, H. J. Kim, D. H. Oh, J. R. Ahn, and S. U. Son, Small **4**, 945 (2008).
[9] J. Sun, R.-S. Deacon, W. Luo, Y. Yuan, X. Liu, H. Xie, Y. Gao, and K. Ishibashi, Commun. Phys. 3, 93(2020).
[10] J. Zhong, J. Yu, L. Cao, C. Zeng, J. Ding, C. Cong, Z. Liu, and Y. Liu, Nano Research **13**, 1780-1786(2020).
[11] B. Radisavljevic, A. Radenovic, J. Brivio, V. Giacometti, and A. Kis, Nat. Nanotechnol **6**, 147 (2011).
[12] Q. H. Wang, K. Kalantar−Zadeh, A. Kis, J. N. Coleman, and M. S. Strano, Nat. Nanotechnol **7**, 699 (2012).
[13] H. Wang, L. Yu, Y. H. Lee, Y. Shi, A. Hsu, M. L. Chin, L. J. Li, M. Dubey, J. Kong, and T. Palacios, Nano. Lett. **12**, 4674 (2012).
[14] A. K. Geim, and I. V. Grigorieva, Nature (London) **499**, 419 (2013).
[15] P. Goli, J. Khan, D. Wickramaratne, R. K. Lake, and A. A. Balandin, Nano Lett **12**, 5941 (2012).
[16] R. Claessen, R. O. Anderson, J. W. Allen, C. G. Olson, C. Janowitz, W. P. Ellis, S. Harm, M. Kalning, R. Manzke, and M. Skibowski, Phys. Rev. Lett. **69**, 808 (1992).
[17] J. W. Allen, G. H. Gweon, R. Claessen, and K. Matho, Journal of Physics and Chemistry of Solids **56**, 1849 (1995).
[18] R. Claessen, R.-O. Anderson, G.-H. Gweon, J.-W. Allen, W.-P. Ellis, C. Janowitz, C.-G. Olson, Z.-X. Shen, V. Eyert, M. Skibowski, K. Friemelt, E. Bucher, and S. Hufner, Phys. Rev. B **54**, 2453 (1996).
[19] L. Perfetti, C. Rojas, A. Reginelli, L. Gavioli, H. Berger, G. Margaritondo, M. Grioni, R. Gaál, L. Forró, and F. Rullier Albenque, Phys. Rev. B **64**, 115102 (2001).
[20] K. Rossnagel, L. Kipp, M. Skibowski, C. Solterbeck, T. Strasser, W. Schattke, D. Voβ, P. Krüger, A. Mazur, and J. Pollmann, Phys. Rev. B **63**, 125104 (2001).
[21] L. Perfetti, C. Rojas, A. Reginelli, L. Gavioli, H. Berger, G. Margaritondo, and M. Grioni, Surface Review and Letters **9**, 1117 (2012).
[22] G. Nicolay, B. Eltner, S. Hfner, F. Reinert, U. Probst, and E. Bucher, Phys. Rev. B **73**, 045116 (2006)
[23] X. F. Tang, Y. X. Duan, F. Y. Wu, S. Y. Liu, C. Zhang, Y. Z. Zhao, J. J. Song, Y. Luo, Q. Y. Wu, J. He, H. Y. Liu, W. Xu, and J. Q. Meng, Phys. Rev. B **99**, 125112 (2019).
[24] M. Hangyo, S. Nakashima, and A. Mitsuishi, Ferroelectrics **52**, 151-159 (1983).
[25] V. Rajaji, U. Dutta, P. C. Sreeparvathy, S. C. Sarma, Y. A. Sorb, B. Joseph, S. Sahoo, S. C. Peter, V. Kanchana, and C. Narayana, Phys. Rev. B **97**, 085107 (2018)
[26] H. N. S. Lee, M. Garcia, H. McKinzie, and A. Wold, Journal of Solid State Chemistry **1**, 190 (1970).
[27] M. Naito, and S. Tanaka, Journal of the Physical Society of Japan **51**, 219 (1982).
[28] P. Chen, W. W. Pai, Y. H. Chan, A. Takayama, C. Z. Xu, A. Karn, S. Hasegawa, M. Y. Chou, S. K. Mo, A. V. Fedorov, and T. C. Chiang, Nat. Commun. **8**, 516 (2017).
[29] S. Fragkos, R. Sant, C. Alvarez, A. Bosak, P. Tsipas, D. Tsoutsou, H. Okuno, G. Renaud, and A. Dimoulas, Advanced Materials Interfaces **6**, 1801850 (2019).
[30] Z. Zhu and Y. Cheng and U. Schwingenschlögl, Phys. Rev. Lett. **110**, 077202 (2013).
[31] Q. Zhang, Y. Cheng, and U. Schwingenschlögl, Phys. Rev. B **88**, 155317 (2013).
[32] M. Zhang, X. Wang, A. Rahman, Q. Zeng, D. Huang, R. Dai, Z. Wang, and Z. Zhang, Appl. Phys. Lett. **112**, 041907 (2018).
[33] U. Dutta, P. S. Malavi, S. Sahoo, B. Joseph, and S. Karmakar, Phys. Rev. B **97**, 060503(R) (2018).
[34] J. Demsar, K. Biljaković, and D. Mihailovic, Phys. Rev. Lett. **83**, 800 (1999).
[35] X. Zhang, H. Y. Song, X. C. Nie, S. B. Liu, Y. Wang, C. Y. Jiang, S. Z. Zhao, G. Chen, J. Q. Meng, Y. X Duan, and H. Y. Liu, Phys. Rev. B **99**, 125141 (2019).
[36] See Supplemental Material for additional data of 1T-TiTe$_2$, which includes Refs.[37, 38].
[37] S. Engelsberg and J. R. Schrieffer, Phys. Rev. **131**, 993(1963).
[38] P. B. Allen, Phys. Rev. Lett. **59**, 1460 (1987).

[39] D. K. G. de Boer, C. F. van Bruggen, G. W. Bus, R. Coehoorn, C. Haas, G. A. Sawatzky, H. W. Myron, D. Norman, and H. Padmore, Phys. Rev. B **29**, 6797 (1984).
[40] N. V. Smith, and M. M. Traum, Phys. Rev. B **11**, 2087 (1975).
[41] H. P. Hughes, C. Webb and P. M. Williams, Journal of Physics C: Solid State Physics **13**, 1125 (1980).
[42] Z. Wang, S. McKeown Walker, A. Tamai, Y. Wang, Z. Ristic, F. Y. Bruno, A. de la Torre, S. Ricc, N. C. Plumb, M. Shi, P. Hlawenka, J. Snchez-Barriga, A. Varykhalov, T. K. Kim, M. Hoesch, P. D. C. King, W. Meevasana, U. Diebold, J. Mesot, B. Moritz, T. P. Devereaux, M. Radovic, and F. Baumberger, Nature Materials **15**, 835(2016).
[43] A. Kerelsky, L. J. McGilly, D. M. Kennes, L. Xian, M. Yankowitz, S. Chen, K. Watanabe, T. Taniguchi, J. Hone, C. Dean, A. Rubio, and A. N. Pasupathy, Nature (London) **572**, 95 (2019).
[44] Y. Jiang, X. Lai, K. Watanabe, T. Taniguchi, K. Haule, J. Mao, and E. Y. Andrei, Nature (London) **573**, 91 (2019).
[45] S. Lisi, X. B. Lu, T. Benschop, T. A. de Jong, P. Stepanov, J. R. Duran, F. Margot, I. Cucchi, E. Cappelli, A. Hunter, A. Tamai, V. Kandyba, A. Giampietri, A. Barinov, J. Jobst, V. Stalman, M. Leeuwenhoek, K. Watanabe, T. Taniguchi, L. Rademaker, S. Jan van der Molen, M. P. Allan, D. K. Efetov, and F. Baumberger, Nat. Phys. **17**, 189(2020).
[46] Y. He, S. D. Chen, Z. X. Li, D. Zhao, D. j. Song, Y. Yoshida, H. Eisaki, T. Wu, X. H. Chen, D. H. Lu, C. Meingast, R. J. Birgeneau, T. P. Devereaux, M. Hashimoto, D. H. Lee, Z. X. Shen, arXiv:2009.10932
[47] M. Schärli, J. Brunner, H. P. Vaterlaus, and F. Levy, J. Phys. C, **16**, 1527 (1983).
[48] F. Schrodi, A. Aperis, P. M. Oppeneer, arXiv:2011.07021
[49] J. Qi, T. Durakiewicz, S.-A. Trugman, J.-X. Zhu, P.-S. Riseborough, R. Baumbach, E.-D. Bauer, K. Gofryk, J.-Q. Meng, J.-J. Joyce, A.-J. Taylor, and R.-P. Prasankumar, Phys. Rev. Lett. **111**, 057402(2013).
[50] Y.-P. Liu, Y.-J. Zhang, J.-J. Dong, H. Lee, Z.-X. Wei, W.-L. Zhang, C.-Y. Chen, H.-Q. Yuan, Y. F. Yang, and J. Qi, Phys. Rev. Lett. **124**, 057404(2020).
[51] X. C. Nie, H. Y. Song, X. Zhang, Y. Wang, Q. Gao, L. Zhao, X.J. Zhou, J. Q. Meng, Y. X. Duan, H. Y. Liu, and S. B. Liu, Physica C: Superconductivity and its Applications **577**, 1353710 (2020).
[52] J. P. Hinton, J. D. Koralek, G. Yu, E. M. Motoyama, Y. M. Lu, A. Vishwanath, M. Greven, and J. Orenstein, Phys. Rev. Lett. **110**, 217002 (2013).
[53] Y. C. Tian, W. H. Zhang, F. S. Li, Y. L. Wu, Q. Wu, F. Sun, G. Y. Zhou, L. Wang, X. C. Ma, Q. K. Xue, and J. M. Zhao, Phys. Rev. Lett. **116**, 107001(2016).
[54] X. C. Nie, H. Y. Song, X. Zhang, P. Gu, S.B. Liu, F. Li, J. Q. Meng, Y. X. Duan, and H. Y. Liu, New Journal of Physics **20**, 033015(2018).
[55] J. Qi, X. Chen, W. Yu, P. Cadden-Zimansky, D. Smirnov, N.-H. Tolk, I. Miotkowski, H. Cao, Y.-P. Chen, Y. Wu, S. Qiao, and Z. Jiang, APPLIED PHYSICS LETTERS **97**, 182102(2010).
[56] V. Iyer, Y.P. Chen, and X. Xu, Phys. Rev. Lett. **121**, 026807 (2018).
[57] Q. Wu, H. X. Zhou, Y. L. Wu, L. L. Hu, S. L. Ni, Y. C. Tian, F. Sun, F. Zhou, X. L. Dong, Z. X. Zhao, and J. M. Zhao, Chin. Phys. Lett. **37**, 097802(2020).
[58] Y. M. Dai, J. Bowlan, H. J. Li, H. Miao, S. F. Wu, W. D. Kong, Y. G. Shi, S. A. Trugman, J. X. Zhu, H. Ding, A. J. Taylor, D. A. Yarotski, and R. P. Prasankumar, Phys. Rev. B **92**, 161104(R) (2015).
[59] Y. M. Sheu, Y. J. Chien, C. Uher, S. Fahy, and D. A. Reis, Phys. Rev. B **87**, 075429 (2013).
[60] M. Balkanski, R. F. Wallis, and E. Haro, Phys. Rev. B **28**, 1928 (1983).
[61] J. Menendez and M. Cardona, Phys. Rev. B **29**, 2051 (1984).
[62] J. Q. Zhou, R. Bianco, L. Monacelli, I. Errea, F. Mauri, and M. Calandra, 2D Materials **7**, 045032(2020).
[63] R.-C. Xiao, W.-J. Lu, D.-F. Shao, J.-Y. Li, M.-J. Wei, H.-Y. Lv, P. Tong, X.-B. Zhu, and Y.-P. Sun, J. Mater. Chem. C **5**, 4167 (2017).
[64] S. Hunsche, K. Wienecke, T. Dekorsy, and H. Kurz, Phys. Rev. Lett. **75**, 1815(1995).
[65] V. Recoules, J. Clerouin, G. Zerah, P. M. Anglade, and S. Mazevet, Phys. Rev. Lett. **96**, 055503(2006).
[66] N. S. Grigoryan, T. Zier, M. E. Garcia, and E. S. Zijlstra, J. Opt. Soc. Am. B **31**, C22–C27(2014).
[67] G. Q. Yan, X. L. Cheng, H. Zhang, Z.Y. Zhu, and D. H. Ren, Phys. Rev. B **93**, 214302(2016).